\documentclass[10pt,aps,showpacs,twocolumn,unsortedaddress,nofootinbib,floatfix]{revtex4-1}
\usepackage[dvipsnames]{xcolor}
\usepackage{graphicx}
\usepackage{bm}
\usepackage{physics}
\usepackage[version=4]{mhchem}
\usepackage{mathtools,mathrsfs,amsfonts,dsfont}
\usepackage{relsize}
\usepackage{scalerel}
\usepackage{todonotes}
\usepackage{ulem}
\usepackage{xcolor}
\usepackage[unicode=true, pdfusetitle, bookmarks=true, bookmarksnumbered=false, bookmarksopen=false, breaklinks=true, pdfborder={0 0 0}, backref=false, colorlinks=true, linkcolor=blue, citecolor=blue, urlcolor=blue]{hyperref}
\usepackage{enumitem}
\usepackage{cancel}
\allowdisplaybreaks
\newcommand{\Ep}{\,\skew{4}{\vu}{\boldsymbol{\phi}}}
\newcommand{\tpsi}{\skew{3}{\tilde}{\psi}}
\newcommand{\tvarphi}{\skew{3}{\tilde}{\varphi}}

\newcommand{\sm}{\kern0.1em}

\begin{document}

\title{Times of Arrival and Gauge Invariance}

\author{Siddhant Das}
\email{Siddhant.Das@physik.uni-muenchen.de}
\author{Markus N\"{o}th}
\email{noeth@math.lmu.de}
\affiliation{Mathematisches Institut, Ludwig-Maximilians-Universitat M\"{u}nchen, Theresienstr. 39, D-80333 M\"{u}nchen, Germany}

\date{December 18, 2020}

\begin{abstract}
We revisit the arguments underlying two well-known arrival-time distributions in quantum mechanics, viz., the Aharonov-Bohm and Kijowski (ABK) distribution, applicable for freely moving particles, and the quantum flux (QF) distribution. An inconsistency in the original axiomatic derivation of Kijowski's result is pointed out, along with an inescapable consequence of the ``negative arrival times'' inherent to this proposal (and generalizations thereof). The ABK free-particle restriction is lifted in a discussion of an explicit arrival-time setup featuring a charged particle moving in a constant magnetic field. A natural generalization of the ABK distribution is in this case shown to be critically gauge-dependent. A direct comparison to the QF distribution, which does not exhibit this flaw, is drawn (its acknowledged drawback concerning the quantum backflow effect notwithstanding).
\end{abstract}

\maketitle
\normalem
\section{Introduction}
The distribution of arrival (or detection) times of a quantum particle amenable to laboratory time-of-flight (TOF) experiments is far from settled, as evidenced by the multitude of inequivalent theoretical predictions suggested in the literature \cite{MUGA,MUGA1,MSP,backwall1}. In a typical theoretical discussion of a TOF experiment, one considers a particle of mass \(\smash{m}\) with a well-localized wave function \(\smash{\psi_0(\vb{x})}\) at time zero, propagating either freely or in specified external potentials. Critical to any such discussion is the probability \(\smash{\Pi(\tau)\,d\tau}\) that the particle's time-of-arrival on a given surface \(\smash{Q}\) is between times \(\smash{\tau}\) and \(\smash{\tau+d\tau}\), subject to the condition
\begin{equation}\label{normal}
    \int_0^{\infty}\!\!\!d\tau~\Pi(\tau)=1.
\end{equation}
For completeness, one could add a ``non-detection probability'' \(\smash{P(\infty)}\) to the left-hand side of \eqref{normal}, accounting for the fraction of experimental runs in which the particle does not intercept \(\smash{Q}\) even as \(\smash{t\to\infty}\).

Most approaches lead to an ideal (or intrinsic) TOF distribution \(\smash{\Pi(\tau)}\), i.e., an {\emph {apparatus-independent}} theoretical prediction, given by some functional of the initial quantum state \(\smash{\ket{\psi_0}}\) and the surface \(\smash{Q}\). In the category of ideal TOF distributions --- insofar as such a description may be warranted --- the Aharonov-Bohm-Kijowski (ABK), and the quantum flux (QF) distributions are exemplary. These distributions have been arrived at from different theoretical viewpoints, making them important benchmarks for comparing with experiments. However, it has long been observed that both predictions practically coincide in the presently accessible far-field or scattering regime, making an experimental test distinguishing the two proposals very challenging.

We provide a critical review of these developments in Sec. \ref{viewpoints}, pointing out an inconsistency in the original axiomatic derivation of Kijowski's distribution. An unexpected consequence of the ``negative arrival times'', characteristic of the ABK proposal and generalizations thereof, is discussed, which casts doubts on the associated time-energy uncertainty relation. The QF distribution is motivated as a special case of the arrival-time distribution in Bohmian mechanics (de Broglie-Bohm or pilot-wave theory) but has also been arrived at from other angles in specific instances. Focusing next on the arrival times of particles subject to external potentials, we are led to the question of gauge invariance. A natural generalization of the ABK formula to scalar potentials, the so-called ``standard arrival-time distribution'', and its direct extension to vector potentials are introduced. For concreteness, we study the arrival-time distributions of a charged particle in a constant magnetic field (Sec. \ref{Bfield}), discovering that the said extension is not gauge-invariant, while the QF distribution is. The former even ceases to be a meaningful probability distribution in some gauges. We conclude in Sec. \ref{Conc}, drawing lessons for future work. In what follows, we take as the units of measurement of mass, length, and time, \(\smash{m}\), \(\smash{\sigma}\) (the width of the wave packet \(\smash{\psi_0}\)), and \(\smash{\sigma^2m/\hbar\sm}\), respectively; formally, this amounts to setting \(\smash{\hbar=m=\sigma=1}\) in every equation.
\section{Theoretical viewpoints}\label{viewpoints}
Work by Aharonov-Bohm \cite{AhBohm} and Paul \cite{HPaul} in the 1960s suggested a TOF distribution starting with the classical arrival-time formula
\begin{equation}\label{clas}
    \tau = \frac{L-z}{p} 
\end{equation}
describing a freely moving particle that at \(\smash{t=0}\) had position \(\smash{z}\) and momentum \(\smash{p}\), and arrived after time \(\smash{\tau}\) at a distant point \(\smash{L}\) on a line\footnote{The correct classical TOF formula valid for an arbitrary initial point \(\smash{(z,p)}\) in phase space is
\begin{equation*}
    \tau = \begin{cases}\,(L-z)/ p, & \text{sgn}\sm p=\text{sgn}\sm(L-z),\\[2.5pt]\,\infty, &\text{otherwise},\end{cases}
\end{equation*}
quantizing which seems nothing short of impossible.\label{fullformula}}. They sought the symmetric quantization \cite[Sec. 5]{MSP}
\begin{equation}\label{ABop}
    \hat{\tau}=L\sm\hat{p}^{\sm-\,1}-\frac{1}{2}\,\big(\hat{p}^{\sm-\,1}\,\hat{z}\,+\,\hat{z}\,\hat{p}^{\sm-\,1}\big),
\end{equation}
promoting \eqref{clas} to the Hilbert space operator \(\smash{\hat{\tau}}\), ostensibly a quantum observable, where \(\smash{\hat{z}}\) and \(\smash{\hat{p}=-\,i\partial/\partial z}\) denote the usual position and momentum operators, respectively, of quantum mechanics. Formally, \eqref{ABop} is canonically conjugate to the free-particle Hamiltonian, i.e.,
\begin{equation}\label{CCR}
    \big[\hat{p}^2,\hat{\tau}\big]=2\sm i\mathds{1},
\end{equation}
although \(\smash{\hat{\tau}}\) is not a self-adjoint operator (i.e., a quantum observable in the sense of Dirac and von Neumann). Nevertheless, its (generalized) eigenfunctions constitute an over-complete set \cite{MugaBad}, defining a positive operator-valued measure (or POVM)\footnote{Quantum observables defined by self-adjoint operators cannot be found for many other experiments as well. The notion of an observable was thus generalized to POVMs \cite[pp. 480--484]{RodiPOVMIntro}, and there are various (inequivalent) suggestions for arrival-time POVMs.}, and in turn the arrival-time distribution \cite{GP,MugaBadBad}
\begin{align}\label{AB1D}
    \Pi_{\text{AB}}(\tau) &= \frac{1}{2\sm\pi}\!\sum_{\alpha\sm=\sm\pm}\left|\sm\int_{-\infty}^{\infty}\!\!\!dp~\theta(\alpha p)\,\sqrt{|p|}\, \tpsi_0(p)\right.\nonumber\\
    &\kern2cm\phantom{\int_{-\infty}^{\infty}}\left.\times\,\exp(-\,\frac{i\tau}{2}\sm p^2+ipL)\right|^2\!\!.
\end{align}
Here, \(\smash{\tpsi_0(p)=\braket{p}{\psi_0}}\) is the momentum representation of the initial state \(\smash{\ket{\psi_0}}\) and \(\smash{\theta(\cdot)}\) is Heaviside's step function. A reformulation of \eqref{AB1D} due to Leavens \cite{LeavensNL,*LeavensNLNL},
\begin{align}\label{leavensformula}
    \Pi_{\text{AB}}(\tau) &= \frac{1}{32\sm\pi}\!\sum_{\alpha\sm=\sm\pm}\left|\int_{-\infty}^{\infty}\!\!\!dz~\frac{1+i\alpha\sm\text{sgn}\sm(z-L)}{|z-L|^{3/2}}\right.\nonumber\\
    &\kern2.5cm\left.\phantom{\int_{-\infty}^{\infty}}\times\, \big[\psi_\tau(z)-\psi_\tau(L)\big]\sm\right|^2\!\!,
\end{align}
involves only the time-dependent position space wave function. Several authors have arrived at (or endorsed) \(\smash{\Pi_{\text{AB}}}\) \cite{GRT,HMM,Savvidou,Seidel,CTOAKij}, most notably Allcock \cite{Allcock2,Allcock3} and Kijowski \cite{kijTOF} (see below). Allcock deduced only the \(\smash{\alpha=+}\) contribution of \eqref{AB1D}, starting from a phenomenological detector model based on a complex absorbing potential (see also \cite{Seidel03}).

Due to the negative eigenvalues of \(\smash{\hat{\tau}}\) (perhaps attributable to the incomplete classical formula, see footnote \ref{fullformula}), \(\smash{\Pi_{\text{AB}}(\tau)}\) is normalized over the interval \(\smash{-\infty<\tau<\infty}\), as opposed to Eq. \eqref{normal}. A typical narrative accompanying these ``negative arrival times'' characteristic of (\ref{AB1D}-\ref{leavensformula}) and generalizations thereof invokes the notion of a quantum state prepared in the infinite past that evolves into \(\smash{\psi_0}\) at time zero (see \cite[p. 155]{LeavOTS}, \cite[p. 4679]{GRT}, or \cite[p. 4337]{MSP} for details). The crux of these arguments is that the negative arrival-time probability should be interpreted as the non-detection probability; accordingly,
\begin{equation}\label{nondet}
    P_{\text{AB}}(\infty) = \int_0^{\infty}\!\!\!d\tau~\Pi_{\text{AB}}(-\sm\tau).
\end{equation}
This then precludes the application of usual quantum formulas for expectation values that inevitably include negative eigenvalue contributions. In particular, taking \(\smash{\matrixel{\psi}{\hat{\tau}}{\psi}}\) as the mean arrival-time and following the work in \cite{Galaponsad} generalized to account for \(\smash{L\neq 0,}\) we find that
\begin{align}\label{expint}
    \matrixel{\psi}{\hat{\tau}}{\psi} &= \frac{i}{4}\int_{\mathbb{R}^2}\!\!dz\, dz^\prime~\big(2\sm L-z-z^\prime\big)\,\text{sgn}\big(z-z^\prime\big)\nonumber\\
    &\kern3.5cm\times\,\psi_0^*\big(z\big)\sm\psi_0^{\phantom{*}}\!\big(z^\prime\big),
\end{align}
hence \(\smash{\expval{\tau}}\) vanishes for any \emph{real} \(\smash{\psi_0(z)}\) \(\smash{\big(\sm\text{sgn}\sm(\cdot)}\) is odd, implying \(\smash{\matrixel{\psi}{\hat{\tau}}{\psi}=-\sm\matrixel{\psi}{\hat{\tau}}{\psi}\!\big)}\), leading to the absurd conclusion that every arrival is \emph{instantaneous} in these cases.

In view of Eq. \eqref{CCR}, one could scrutinize some version of the time-energy uncertainty relation \cite[Ch. 3]{MUGA}, the main concern of Aharonov-Bohm \cite{AhBohm} and Kijowski \cite{kijTOF} among others. However, the distribution (\ref{AB1D}-\ref{leavensformula}) decays too slowly to have finite first and second moments\footnote{This by no means invalidates a TOF distribution so long as it is normalizable \`a la Eq. \eqref{normal}, which merely forces \(\smash{\Pi(\tau)}\) to decay faster than \(\smash{1/\tau}\) as \(\smash{\tau\to\infty}\). \label{momfn}}, unless \(\smash{\ket{\psi_0}}\) satisfies
\begin{equation}\label{domAB}
    \lim_{p\to0} p^{\sm-\,3/2}\,\tpsi_0(p) = 0.
\end{equation}
As a result, one typically restricts the domain of the operator \(\smash{\hat{\tau}}\) to the subspace of wave functions fulfilling this condition \cite{MugaBadBad,MugaBad}. But this leaves out, e.g., Gaussian wave packets \(\smash{\propto \exp(-\,\alpha\sm z^2+i\beta z)}\) that violate \eqref{domAB} producing an infinite \(\smash{\Delta\tau}\), being otherwise perfectly reasonable states amenable to present-day experiments \cite{coolingA,coolingB,Ferdicooling}. Secondly, even when the moments are finite, they may no longer satisfy the desired (Robertson-Schr\"odinger) uncertainty inequality, once the contribution of the ``negative arrival times'' is discarded from expectation value integrals \cite[p. 190]{SV}. Therefore, substantiating the time-energy uncertainty relation via arrival-time operators fulfilling \eqref{CCR}, of which \eqref{ABop} is but one example \cite{GRT,Galaponantithesis,Galaponsad,CTOA,Recami,Goto1,Goto2,Rosenbaum,academic}, seems academic at best. This does little to generate enthusiasm for cutting-edge TOF experiments. Quite to the contrary.

In 1974 Kijowski \cite{kijTOF} rediscovered \eqref{AB1D} (rather, a three-dimensional version of it applicable to arrivals on a plane) by imposing a set of plausible axioms he believed a TOF distribution should satisfy. He initially considered freely moving particles, for which
\begin{equation}\label{timevol}
    \tpsi_t(\vb{p})=\tpsi_0(\vb{p})\sm \exp(-\,\frac{it}{2}\sm p^2),
\end{equation}
prepared in special initial states that --- in momentum representation --- vanish for \(\smash{p_z\le0}\), the so-called ``right moving wave functions''. For arrivals on the plane \(\smash{z=L}\), he parameterized the arrival-time probability density as
\begin{equation}\label{PiKij}
    \Pi_{\text{Kij}}(\tau) = F\big(e^{ip_z L}\sm\tpsi_\tau\big),
\end{equation}
where \(\smash{F(\cdot)}\) is a quadratic form \cite{kijTOF,BohmbeatsKij}, itself defining the arrival-time probabilities for the \(\smash{xy}\)-plane (or \(\smash{z=0}\)). He required the following properties of \(\smash{F}\):
\begin{enumerate}[label=(\roman*),leftmargin=3cm,align=left]
\item $\displaystyle F(\psi)\ge0,$\label{i}
\item $\displaystyle F(\psi^*)=F(\psi),$\label{ii}
\item $\displaystyle F\big(\hat{U}\psi\big)=F(\psi),$\label{iii}
\item $\displaystyle\int_{-\infty}^{\infty}\!\!\!dt~F(\psi_t)=\braket{\psi_0}{\psi_0}=1,$\label{iv}
\end{enumerate}
and
\begin{enumerate}[resume,label=(\roman*),leftmargin=3cm,align=left]
\item $\displaystyle \int_{-\infty}^{\infty}\!\!\!dt~t^2\,F(\psi_t)<\infty,$\label{v}
\end{enumerate}
drawing upon his analysis of the analogous classical arrival-time problem. In \ref{iii}, \(\smash{\hat{U}}\) denotes the (unitary) representation of any Galilei transformation that leaves the plane of arrival invariant, e.g., translations parallel to the \(\smash{xy}\)-plane. Although infinitely many quadratic forms fulfill \ref{i}-\ref{v}, he noticed that\footnote{The different prefactor in \cite[Eq. (9)]{kijTOF} is due to a different Fourier transform convention.}
\begin{equation}\label{F0}
    F_0(\psi) = \frac{1}{2\sm\pi}\int_{\mathbb{R}^2}\!\!dp_x\,dp_y\left|\sm\int_0^{\infty}\!\!\!dp_z~\sqrt{p_z}~\tpsi(\vb{p})\sm\right|^2
\end{equation}
was \emph{unique}, in that  
\begin{align}
    \int_{-\infty}^{\infty}\!\!\!dt~t\,F(\psi_t) &= \int_{-\infty}^{\infty}\!\!\!dt~t\,F_0(\psi_t),\label{result1}\\&\kern-4cm\text{and}\nonumber\\
    \int_{-\infty}^{\infty}\!\!\!dt~t^2\,F(\psi_t) &\ge \int_{-\infty}^{\infty}\!\!\!dt~t^2\,F_0(\psi_t),\label{result2}
\end{align}
given any admissible \(\smash{F}\). As a result, he found it ``quite reasonable to consider \(\smash{F_0(\psi_t)}\) the true probability density of passing through \(\smash{Q}\) at \(\smash{t}\)'' \cite[p. 368]{kijTOF}. Concerning ``left moving'' wave functions (those vanishing for \(\smash{p_z\ge0}\)), he argued that the solution is analogous to \eqref{F0}, except \(\smash{\sqrt{p_z}\to\!\sqrt{-\sm p_z}\sm}\)  with the momentum integration being performed over \(\smash{p_z\!<\kern-0.1em 0}\). For generic wave functions that are neither ``left'' nor ``right moving'', he proposed
\begin{align}\label{F0full}
    &F_0(\psi)=\frac{1}{2\sm\pi}\!\sum_{\alpha\sm=\sm\pm}\int_{\mathbb{R}^2}\!\!dp_x\,dp_y\left|\sm\int_{-\infty}^{\infty}\!\!\!dp_z\,\theta(\alpha p_z)\right.\nonumber\\
    &\kern2cm\left.\phantom{\sm\int_{-\infty}^{\infty}\!\!\!dp_z\,\theta(\alpha p_z)}\times\,\sqrt{|p_z|}~\tpsi(\vb{p})\sm\right|^2\!\!.
\end{align}
The complete TOF distribution is thus specified by Eqs. (\ref{timevol}-\ref{PiKij}) and \eqref{F0full}, which reproduces \(\smash{\Pi_{\text{AB}}(\tau)}\), Eq. \eqref{AB1D}, in the appropriate one-dimensional context.

It follows that Kijowski's approach also admits ``negative arrival times'' (in fact, right from the beginning via axioms \ref{iv} and \ref{v}), for which no explanation is offered in \cite{kijTOF}, nor in the follow-up paper \cite{Kij2}. As noted before, if these are taken literally, one faces the befuddling consequence that the predicted mean arrival-time vanishes for any real-valued (position space) wave function prepared at time zero\footnote{These wave functions fulfill, in momentum space, \(\smash{\tpsi_0^{\phantom{*}}\!(-\sm\vb{p})=\tpsi_0^*(\vb{p})}\), making \eqref{PiKij} a symmetric function of \(\smash{\tau}\) in view of Eq. \eqref{timevol} and axiom \ref{ii}, from which follows the claimed consequence.}. The same holds for any other \(\smash{F\kern-0.1em}\), given Eq. \eqref{result1}.

In addition, time integrals occurring in axiom \ref{v} and the results (\ref{result1}-\ref{result2}) are assumed to be well-defined for all wave functions. That this assumption is questionable is readily seen: consider, e.g., the ``right moving'' wave function
\begin{equation}
    \tvarphi_0(\vb{p})=\frac{\!\sqrt{2}}{\pi^{3/4}}~\theta(p_z)\,\sm e^{-\sm p^2/2},
\end{equation}
which yields via Eqs. \eqref{timevol} and \eqref{F0}
\begin{equation}\label{ce2}
   F_0(\varphi_t) = \frac{F_0(\varphi_0)}{(1+t^2)^{3/4}},
\end{equation}
where \(\smash{F_0(\varphi_0)}\) is an (irrelevant) multiplicative constant. Note that \(\smash{t^2F_0(\varphi_t)\sim\kern-0.1em\sqrt{|t|\sm}}\), as \(\smash{t\to\pm\,\infty}\), thus contradicting axiom \ref{v} --- the very input that led to the unique identification of \(\smash{F_0}\). Some other axiomatic approaches \`a la Kijowski \cite{HM,HMM,WernerScreen} also make use of axiom \ref{v}, but focus on a select subset of wave functions.

We emphasize that the existence, or lack thereof, of moments, is rather superfluous for the physical relevance of any arrival-time distribution (cf. footnote \ref{momfn}). Recall, e.g., the statistics of first-arrival times of a Brownian particle reaching some specified position distant from its initial location, described by the well-known L\'evy distribution having \emph{no} finite moments \cite[Ch. 10]{Balaki}. In this respect, \(\smash{\Pi_{\text{Kij}}(\tau)}\) is a perfectly reasonable proposal. However, the question remains of whether a more principled axiomatic derivation of Kijowski's results, either restricting attention to some reasonable subspace of \(\smash{L^2\big(\mathbb{R}^3\big)}\) (albeit compromising on the POVM structure), or forsaking axiom \ref{v} altogether could be given. Preferably, one that evades the problematic ``negative arrival times''.

Freely moving particles have hitherto been the center of our focus, the natural next step is to describe the arrival times of particles moving in specified potentials and/or arriving on nonplanar surfaces. Even individually, these tasks have proven to be anything but simple, giving rise to a myriad of inequivalent generalizations (primarily restricted to one spatial dimension) of the ABK free-motion formulae
\cite{HM,HMM,Leon,Baute,Baute1,Galaponantithesis,Galaponsad,Goto1,Goto2}.
A comprehensive account of all these proposals, including a novel experimental setup for distinguishing one from another and, in particular, from the quantum flux (Eqs. (\ref{qf3d}-\ref{current}) below), appears in \cite{backwall1}.

For concreteness, we focus on a particularly straightforward generalization of Kijowski's distribution \cite{Baute,Baute1}, \cite[Sec. 10.6]{MUGA}, referred to as the ``standard arrival-time distribution'' by some authors:
\begin{align}\label{pistd}
    \Pi_{\text{STD}}(\tau) &= F_0\big(e^{ip_z L}\sm\tpsi_\tau\big),
\end{align}
with \(\smash{F_0}\) defined in \eqref{F0}. It is formally identical to Kijowski's distribution, Eq. \eqref{PiKij}, but its scope has now been enlarged\footnote{Kijowski's own view \cite{kijTOF} that his ``construction is set up for free particles [\(\,\dotsi\sm\)] and cannot be generalized for the non-free wave equation'' notwithstanding.} by letting \begin{equation}\label{genevol}
    \tpsi_t(\vb{p}) = \matrixel{\vb{p}}{\sm\exp(-\,it\hat{H})\sm}{\psi_t}\kern-0.1em,
\end{equation}
where \(\smash{\hat{H}}\) may contain an arbitrary scalar potential \(\smash{V(\vb{x})}\). However, there is a caveat regarding \eqref{pistd}: ``in general it is not normalized, and it may be unnormalizable'' \cite{Baute}.\footnote{To see this explicitly, consider a particle moving in the one-dimensional potential well \(\smash{-\,\delta(z)}\) with \(\smash{t=0}\) wave function \(\smash{\psi_0(z)=\exp(-\,|z|)}\). Since this is a bound state of the potential with energy eigenvalue \(\smash{-\,1/2}\), we have \(\smash{\psi_t(z)=\psi_0(z)\,\exp(it/2)}\). It follows that \(\smash{\Pi_{\text{STD}}(\tau) = F_0\big(e^{ip_zL}\psi_0\big),}\) which is a \emph{constant} independent of \(\smash{\tau}\), hence unnormalizable. A more realistic example might be the superposition \(\smash{\propto\,\exp(-\,\sm|z|)+\exp(-\,2\sm|z|+iqz)}\), where the second term propagates towards \(\smash{z=L}\) with a group velocity \(\smash{q}\). The exact solution of schr\"odinger's equation in this case is known \cite[Sec. IV]{delta}, from which one finds that the ``standard distribution'' is again unnormalizable, irrespective of \(\smash{q}\).\label{normalization}} To our knowledge, there is no underlying derivation of \eqref{pistd} delineating its scope to just scalar potentials, but as we show later it should {\emph {not}} be used for vector potentials (cf. Sec. \ref{Bfield}). The ``standard distribution'' has attracted much criticism, see \cite{Mielnik,LeavensNL,LeavensNLNL,LeavOTS}, to which the pertinent response has been the claim that ``the paradoxical aspects dissolve away from the point of view of the standard interpretation of quantum mechanics'' \cite{Mugareply} (see, however, \cite{LeavensReply}). 

Yet another TOF distribution applicable for motion in generic potentials, but notably different from the aforementioned proposals, is the QF distribution
\begin{equation}\label{qf3d}
    \Pi_{\text{QF}}(\tau) = \int_{Q}\,\vb{J}(\vb{x},\tau)\cdot d\vb{s},
\end{equation}
where \(\smash{Q}\) is a (not necessarily planar or infinite) arrival surface in \(\smash{\mathbb{R}^3}\), \(\smash{\vb{J}}\) is the quantum flux (or probability current) density associated with Schr\"odinger's equation, and \(\smash{d\vb{s}}\) the surface measure at \(\smash{\vb{x}}\). The most general form of \(\smash{\vb{J}}\) applicable for, say, a spin-0 particle\footnote{The probability current for particles of spin \(\smash{\ge1/2}\) has additional contributions, see \cite{Mike,Wardunique} and references therein. But the QF distribution, insofar as it is meaningful, continues to assume the same form as \eqref{qf3d}.\label{spinFN}} of mass \(\smash{m}\) (\(\smash{=1}\) in our units) and charge \(\smash{q}\), moving in the presence of specified electromagnetic potentials \(\smash{V,\,\vb{A}}\), reads
\begin{equation}\label{current}
\vb{J}(\vb{x},t) = \text{Im}\sm\big[\psi_t^*\,\grad\psi_t^{\phantom{*}}\!\big]\sm-\sm q\sm\vb{A}(\vb{x},t)\sm|\psi_t|^2,
\end{equation}
where \(\smash{\psi_t(\vb{x})}\) solves the (minimally coupled) Schr\"odinger equation, Eq. \eqref{mincoup} below. To begin with, Eq. \eqref{qf3d} has the correct physical dimension of an arrival-time distribution, and has been arrived at in various formulations of quantum mechanics, for example, Bohmian mechanics \cite{DDGZ,DDGZ96,Leav,Leav98,Grubl,Kreidl,Vona1} and, for freely moving particles in one-dimension\footnote{One-dimensional derivations of \(\smash{\Pi_{\text{QF}}}\), though illustrative, are not conclusive, since in this setting the vectorial character of \eqref{current} is grossly compromised, and implies a reduction of the arrival surface \(\smash{Q}\) to a single point on a line.}, by both the decoherent-histories formulation of quantum mechanics \cite{histories,histories1,Yearsley,Hutem} and the analysis of specific measurement models \cite{Allcock2,currentA,Mugaphoton,MugaReal,Yearsley2011} with the notable exception of \cite{3DMuga} (see also \cite[Sec. 2]{tunnelingcurrent,Ricami}). Eq. \eqref{qf3d} would be a natural guess for an arrival-time distribution from the perspective of scattering theory as well \cite[p. 912]{DDGZ,DDGZ96,CohenT}, however, for reasons discussed below it is not satisfactory for all wave functions. 

First, we sketch the arguably simplest derivation of \eqref{qf3d} via Bohmian mechanics (de Broglie-Bohm or pilot-wave theory), which offers a distinct conceptual advantage in TOF studies owing to the well-defined concepts of point particles and trajectories embedded in it. It is a quantum theory of particles in motion, where the trajectories are determined by the wave function. The Bohmian trajectories of a charged spin-0 particle satisfy the first order differential equation (or guiding equation) \cite{Bohm1,*Bohm2,DurrTeufel}
\begin{equation}\label{gEq}
\dot{\vb{X}}_t + q\sm\vb{A}(\vb{X}_t,t) = \left.\grad S_t(\vb{x})\sm\right|_{\vb{x}=\vb{X}_t}\!,
\end{equation}
with initial condition \(\smash{\vb{X}_0}\). Here, \(\smash{\vb{X}_t\in\mathbb{R}^3}\) is the position of the particle at time \(\smash{t}\) and \(\smash{S_t(\vb{x})}\) is the (real-valued) phase of the complex wave function, defined by its polar decomposition
\begin{equation}\label{polar}
    \psi_t(\vb{x})=\sqrt{\rho_t(\vb{x})}\, e^{i S_t(\vb{x})}\sm,
\end{equation}
which evolves as in standard quantum mechanics according to Schr\"odinger's equation. The equation of motion \eqref{gEq} is the simplest that is Galilean, time-reversal, and gauge-invariant \cite[p. 147]{DurrTeufel}. Given a surface \(\smash{Q}\), the arrival-time (first passage time, or hitting time) of an individual trajectory is thus (implicitly) defined as a function of the initial condition \(\smash{\vb{X}_0}\). But since the initial particle positions realized in a sequence of experiments are random, the arrival times obtained in different runs of a TOF experiment (employing the same initial wave function) are random, as are the positions where the particle strikes the surface \(\smash{Q}\). 

The statistics of arrival times is thus determined by the distribution of \(\smash{\vb{X}_0}\), which in Bohmian mechanics is given by \(\smash{|\psi_0|^2 = \rho_0}\) \cite{DGZBorn,BornRule}, also known as the quantum equilibrium distribution. A notable consequence of this, attributable to the very form of the guidance equation \eqref{gEq}, is that the position of a Bohmian particle remains \(\smash{|\psi_t|^2=\rho_t}\)--distributed at any later time \(\smash{t}\) (this property is called equivariance). Typically, the Bohmian TOF distribution must be computed numerically, but in situations where the trajectories intersect \(\smash{Q}\) \emph{at most} once, it reduces to the QF distribution \eqref{qf3d}, discounting the microscopic details of the particle trajectories (see also \cite{without}).

To see this, one notes that the single crossing condition is mathematically expressible as
\begin{equation}\label{preCPC}
    \vb{v}(\vb{x},t)\cdot \hat{\vb{n}}\ge0,~\,\forall\, \vb{x}\in Q,~\,\text{and}~\,\forall\sm t>0,
\end{equation}
where \(\smash{\vb{v}}\) denotes the Bohmian velocity field, the integral curves of which are the particle trajectories, and \(\smash{\hat{\vb{n}}}\) is the unit normal vector at \(\smash{\vb{x}\in Q}\) directed along the flow. Now, consider an area element \(\smash{d\vb{s}}\) centered at some \(\smash{\vb{x}^\prime\in Q}\). The probability of crossing \(\smash{d\vb{s}}\) between times \(\smash{\tau}\) and \(\smash{\tau+d\tau}\) is given by the probability of finding the particle in the volume element \(\smash{(\vb{v}(\vb{x}^\prime,\tau)\cdot d\vb{s})\,d\tau}\) at time \(\smash{\tau}\) (here, we are following Boltzmann's cylinder argument from statistical mechanics). By virtue of equivariance, this crossing probability equals \(\smash{\rho_\tau(\vb{x}^\prime)\,(\vb{v}(\vb{x}^\prime,\tau)\cdot d\vb{s})\,d\tau}\), and via Eq. \eqref{gEq}, we have \(\smash{\vb{v}(\vb{x},t)=\grad S_t(\vb{x})-q\vb{A}(\vb{x},t)}\) explicitly. Now, using \eqref{polar}, \(\smash{\psi^*_t\sm\grad\psi^{\phantom{*}}_t=(1/2)\sm\grad\rho_t+i\sm\rho_t\grad S_t}\), which yields, in view of Eq. \eqref{current}, \(\smash{(\vb{J}(\vb{x},\tau)\cdot d\vb{s})\,d\tau}\) for the differential crossing probability. Finally, integrating the same over all \(\smash{\vb{x}^\prime\in Q}\), we arrive at \(\smash{\Pi_{\text{QF}}(\tau)\,d\tau}\). Note well that some trajectories may not intersect \(\smash{Q}\) even once, contributing to a non-zero non-detection probability \(\smash{P(\infty)}\) (given by the \(\smash{\rho_0}\) measure of their initial conditions), but otherwise \eqref{qf3d} is always normalized as per Eq. \eqref{normal}.

That said, the QF distribution \eqref{qf3d} is, as is well-known, not positive definite for every wave function \cite{Backflow}. When negative, it ceases to be a meaningful probability density. From the preceding Bohmian derivation one also expects that \(\smash{\Pi_{\text{QF}}}\) would not be a generally valid arrival-time distribution. In particular, it is reasonable only as long as \eqref{preCPC} holds, which is equivalent to the current positivity (or ``no backflow'') condition \cite{DDGZ,*DDGZ96}:
\begin{equation}\label{CPC}
    \forall\sm t>0\quad\text{and}\quad\forall\sm\vb{x}\,\in\,Q,\quad \vb{J}(\vb{x},t)\cdot d\vb{s}\sm\ge\sm0.
\end{equation}
Even though \eqref{CPC} can be violated in principle, see \cite{BF,BF2,BF3,GoussevPRR}, stable backflow situations are, as it happens, very difficult to set up experimentally. This is because, on the one hand, the flux density of freely evolving wave functions becomes approximately radial at large distances from the support of the initial wave function \(\smash{\psi_0}\) \cite{DDGZ,*DDGZ96}: \begin{equation}\label{gr8}
    \vb{J}(\vb{x},t)\approx \frac{\vb{x}}{t^4}\!\left|\tpsi_0\!\left(\frac{\vb{x}}{t}\right)\right|^2\!,
\end{equation}
granting \eqref{CPC} for just about any surface \(\smash{Q}\) placed in the far-field (as is typical of scattering experiments). On the other hand, in a free one-dimensional setting there exist strict limits on the probability transported back across the arrival point \cite{Backflow,BF0}. Moreover, asymptotic current positivity is not limited to free evolution, manifesting even in the presence of various short and long range potentials (see \cite[p. 123]{DDST} and references therein). Probability current backflow is therefore more readily seen in the near-field regime. Note, however, that near-field detections demand acute time resolution equipment, as the particle arrives almost instantaneously at the detectors\footnote{Keeping this in mind, we have recently analyzed a novel experimental setup involving a spin-1/2 particle moving within a cylindrical waveguide \cite{DD,exotic1}, finding significant quantum backflow at arbitrary distances along the waveguide at any given time (cf. footnote \ref{spinFN}).}.

While measurement outcomes in quantum mechanics are described by POVMs, the QF distribution \eqref{qf3d} does not follow from one \cite{Vona1}, although it is close to one in the scattering regime. In particular, as noted in the introduction, the QF and ABK (or ``standard'') distributions\footnote{The ``standard distribution'' is not associated with a POVM either, being unnormalizable for certain wave functions (cf. footnote \ref{normalization}).} practically coincide in the far-field regime accessible to present-day experiments. This has generated some interest in studying the minutest of differences between these proposals \cite{BohmbeatsKij,LeavOTS,LeavensNL,backwall1,Delgado}, wherein the significance of the backflow effect has been persistently emphasized. For free particles, the differences are indeed negligible, since 
\begin{equation}
    \Pi_{\text{Kij}}(\tau) \approx \frac{L}{\tau^4}\kern-0.1em\left|\tpsi_0\!\left(\frac{L}{\tau}\right)\right|^2\!,
\end{equation}
once \(\smash{L,\tau\!\gg\!1}\) with \(\smash{L/\tau\sim\mathcal{O}(1)}\) (see \cite{backwall1} for details), thereby agreeing with \(\smash{\Pi_{\text{QF}}(\tau)}\) via Eq. \eqref{gr8}. In a few examples containing simple interaction potentials, we found good agreement between the QF and the ``standard distribution'', although a general argument is far from available.
\section{Arrival time statistics in a constant magnetic field}\label{Bfield}
In this section we consider a direct extension of the ``standard distribution'' to vector potentials, letting \(\smash{\hat{H}}\) be the minimally coupled Hamiltonian in Eq. \eqref{genevol}. To illustrate its consequences, we describe next a simple arrival-time experiment involving a spin-0 particle of mass \(\smash{m}\) and charge \(\smash{q}\), moving in a constant magnetic field
\begin{equation}\label{Bz}
    \vb{B}(\vb{x}) = 2\sm B_0\,\hat{\vb{z}},
\end{equation}
directed along the \(\smash{z}\)-axis of a right-handed coordinate system. The particle is prepared in a suitable wave function at time zero, \(\smash{\psi_0(\vb{x})}\), and arrival times are monitored at a distant plane \(\smash{z=L}\). As a pretext for the relevance of this problem outside its present context, we mention the TOF measurements of \emph{single} charged particles moving within a Penning trap in mass spectrometry applications, and that of electrons emitted from quantum-Hall edge states \cite[p. 9]{QHallExp,QHallTh}. In what follows, we employ cylindrical polar coordinates \(\smash{\vb{x}=(r,\phi,z)}\), considering two vector potentials
\begin{equation}
    \vb{A}(\vb{x}) = B_0\sm r\Ep,\quad {\text{and}} \quad \vb{A}^\prime(\vb{x}) = \vb{A}(\vb{x})-\eta\,z\,\hat{\vb{z}},
\end{equation}
that yield \eqref{Bz}, i.e., \(\smash{\curl\vb{A}=\curl\vb{A}^\prime=\vb{B}}\). Here, \(\smash{\eta}\) is a free (real) parameter that can be changed without altering the physical magnetic field under consideration. The vector potential \(\smash{\vb{A}}\) corresponds to what is often called the symmetric gauge, and satisfies the Coulomb gauge condition \(\smash{\div\sm\vb{A}=0}\). The vector potentials are related by the gauge transformation
\begin{equation}\label{gaugetrans}
    \vb{A}^\prime = \vb{A} - \grad\big(\eta\sm z^2/2\big).
\end{equation}
The minimally coupled Schr\"odinger equation,
\begin{equation}\label{mincoup}
    i\hbar\,\frac{\partial\psi_t(\vb{x})}{\partial\sm t} = \frac{1}{2\sm m}\sm\big(\kern-0.1em-i\hbar\sm\grad-q\vb{A}\big)^2\psi_t(\vb{x}),
\end{equation}
assumes the form --- adopting as the units of mass, length and time \(\smash{m}\), \(\smash{\sqrt{\hbar\sm/ qB_0}}\), and \(\smash{m/ qB_0}\), respectively ---
\begin{equation}
    i\frac{\partial\psi_t(\vb{x})}{\partial t}=\left(-\,\frac{1}{2}\sm\laplacian+i\frac{\partial}{\partial\phi}+\frac{r^2}{2}\right)\psi_t(\vb{x}).
\end{equation}
The solution of this equation with a Gaussian initial condition, say, \(\smash{\psi_0(\vb{x})=\pi^{-\,3/4}\,e^{-\,\frac{1}{2}\sm(\sm r^2+z^2\sm)}}\), is given by
\begin{equation}\label{psit}
    \psi_t(\vb{x}) = \frac{e^{-\,it}}{\pi^{3/4}\sqrt{1+it}}\sm \exp[-\,\frac{\sm r^2}{2}-\frac{z^2}{2\sm(1+it)}].
\end{equation}
In view of Eq. \eqref{gaugetrans}, the same wave function solution in the \(\smash{\vb{A}^\prime}\) gauge, is
\begin{equation}
    \psi_t^\prime(\vb{x}) = \psi_t(\vb{x})\,e^{-\,i\eta\sm z^2/2}.
\end{equation}
Note that both wave packets spread dispersively in the direction of (and opposite to) the magnetic field, leading to detection events at the distant plane \(\smash{z=L}\), while in the transverse directions no spreading is observed, thanks to an effective harmonic confinement \(\smash{\propto\sm r^2}\) induced by the magnetic field \eqref{Bz}.

The ``standard arrival-time distribution'' for the primed wave function can be evaluated as follows: First, in view of \eqref{genevol}, we note
\begin{align}\label{psitp}
  \tpsi_{\!t}^\prime(\vb{p})  = \frac{e^{-\,it}}{\pi^{3/4}}\sm\sqrt{\frac{\sigma(t)}{1+it}}\,\exp[-\,\frac{1}{2}\sm\Big(p_x^2+p_y^2+\sigma(t)\sm p_z^2\Big)],
\end{align}
where
\begin{equation}
    \frac{1}{\sigma(t)} = \frac{1}{1+it}+i\eta.
\end{equation}
Incorporating the same into \eqref{pistd}, and performing the Gaussian integrals over \(\smash{p_x}\) and \(\smash{p_y}\), we are left with
\begin{align}
    \Pi_{\text{STD}}(\tau) &= \frac{|\sigma(\tau)|}{2\sqrt{\pi^3(1+\tau^2)}}\sum_{\alpha=\pm}\left|\int_{-\infty}^{\infty}\!\!\!dp_z~\theta(\alpha p_z)\sqrt{|p_z|}\right.\nonumber\\
    &\kern1.5cm\times\left.\exp(-\,\frac{\sigma(\tau)}{2}\sm p_z^2+ip_zL)\right|^2\!\!.
\end{align}
Separating this integral into positive and negative \(\smash{p_z}\) contributions and letting \(\smash{p_z\to-p_z}\) in the latter, yields via \cite[Eq. 3.462.1]{GH}:
\begin{align}\label{finalstd}
    \Pi_{\text{STD}}(\tau)&=\frac{1}{8\sqrt{\pi\sm(1+\tau^2)\sm|\sigma(\tau)|}}\, \exp(-\,\frac{\text{Re}\sm[\sigma(\tau)]}{2\,|\sigma(\tau)|^2}\sm L^2)\nonumber\\[3pt]
    &\kern1.5cm\times\!\sum_{\alpha=\pm}\left|D_{-3/2}\!\left(i\alpha L/\kern-0.1em\sqrt{\sigma(\tau)}\sm\right)\right|^2\!\!.
\end{align}
Here, \(\smash{D_\nu(\cdot)}\) denotes the parabolic cylinder function of order \(\smash{\nu}\) (reducing to the familiar Hermite polynomial for \(\smash{\nu=0,\sm1,\sm2\sm\dots}\) \cite[Eq. 9.253]{GH}). The ``standard distribution'' for the unprimed wave function is obtained by setting \(\smash{\eta=0}\) in the above.

\begin{figure}
    \centering
    \includegraphics[trim=0 1mm 3mm 3mm, clip, width=\columnwidth]{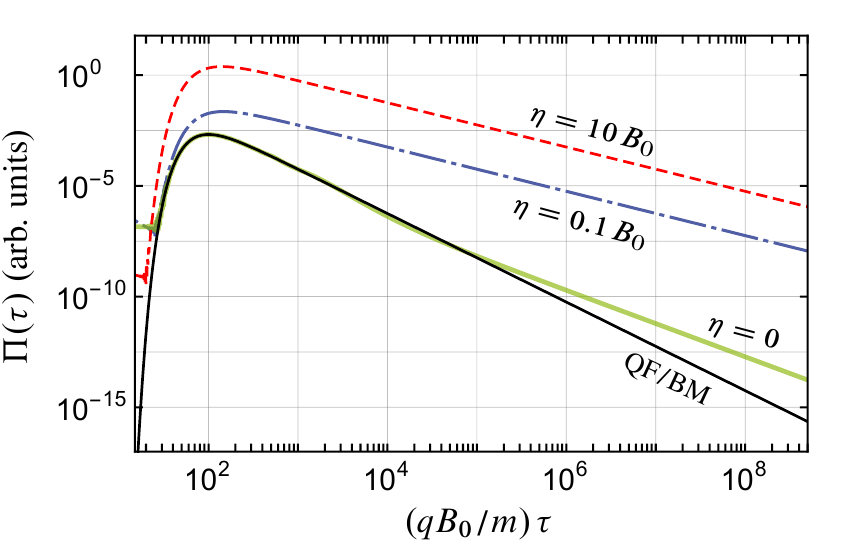}
    \caption{Arrival time distributions \(\smash{\Pi_{\text{QF\,/\,BM}}(\tau)}\) and \(\smash{\Pi_{\text{STD}}(\tau)}\) versus (dimensionless) arrival-time \(\smash{(qB_0/m)\,\tau}\) for \(\smash{L=100\sqrt{\hbar/qB_0}}\) and select values of \(\smash{\eta}\) annotated in the figure. The QF\sm/\sm BM distribution is gauge-invariant, hence independent of \(\smash{\eta}\).}
    \label{fig}
\end{figure}

Since \eqref{finalstd} depends nontrivially on \(\smash{\eta}\) (see Fig. \ref{fig}), and each value of \(\smash{\eta}\) defines the same physical magnetic field \eqref{Bz}, the ``standard distribution'' is not gauge-invariant. Furthermore, for \(\smash{\eta}\) non-zero,
\begin{align}
    \Pi_{\text{STD}}(\tau)&\sim \sqrt{|\eta|\sm}\sm\Big(0.19\,+\,\mathcal{O}\big(\eta\sm L^2\big)\sm+~\dotsi\Big)\,\frac{e^{\sm-\,L^2/2}}{\tau},
\end{align}
as \(\smash{\tau\to\infty}\). Consequently, \eqref{finalstd} is unnormalizable, hence cannot be a physical probability distribution. At this point, one may be tempted to set \(\smash{B_0=0}\), whereby the magnetic field \eqref{Bz} vanishes and the particle is moving freely (in particular, both \(\smash{\vb{B}=\curl\vb{A}^\prime=\vb{0}}\) and \(\smash{\vb{E}=-\,\partial\vb{A}^\prime/\partial t =\vb{0}}\)). In this case, \(\smash{\Pi_{\text{STD}}}\) (which should reproduce \(\smash{\Pi_{\text{Kij}}}\) for freely moving particles) continues to be \(\smash{\eta}\) dependent.

On the other hand, the QF distribution \eqref{qf3d} is given by
\begin{equation}\label{atpresent}
    \Pi_{\text{QF}}(\tau) = \int_0^{2\pi}\!\!\!d\phi\!\int_0^{\infty}\!\!\!dr~r\sm J_z(r,\phi,L,\tau)
\end{equation}
for the case at hand, where 
\begin{equation}\label{Jz}
    J_z(\vb{x},t) = \frac{tz}{\big[\pi\sm (1+t^2)\big]^{3/2}}\, \exp(-\,r^2-\frac{z^2}{1+t^2})
\end{equation}
is the \(\smash{z}\)-component of the (gauge-invariant) probability current density \eqref{current} (the same for either wave function, i.e., for any value of \(\smash{\eta}\)). In particular, \(\smash{J_z>0}\) at \(\smash{z=L}\), i.e., the current positivity condition \eqref{CPC} holds, consequently \eqref{atpresent} would be a meaningful arrival-time distribution. Evaluating \eqref{atpresent}, we obtain
\begin{equation}\label{qfabove}
    \Pi_{\text{QF}}(\tau) = \frac{\tau L}{\sqrt{\pi}}\sm\frac{e^{-\,\frac{L^2}{1+\tau^2}}}{(1+\tau^2)^{3/2}},
\end{equation}
which approaches \(\smash{L/\big(\sqrt{\pi}\sm\tau^2\big)}\), as \(\smash{\tau\to\infty}\). Therefore, it is normalizable (cf. footnote \ref{momfn}), although \eqref{qfabove} is not normalized to unity on \(\smash{0<\tau<\infty}\). This shouldn't come as a surprise, given the symmetry of the wave function \eqref{psit} (or \eqref{psitp}) about \(\smash{z=0}\), implying that the particle should not be detected at \(\smash{z=L}\) in roughly half of the experimental runs, hence \(\smash{P(\infty)\approx 1/2}\). The QF distribution is graphed along with the ``standard distribution'' \eqref{finalstd} in Fig. \ref{fig} for three different values of \(\smash{\eta}\).

It is also informative to directly compute the Bohmian arrival-time distribution for this case. First, the guiding equation \eqref{gEq} implies for either vector potential the component equations
\begin{equation}
    \dot{R}_t=0,\quad \dot{\Phi}_t=1,\quad \dot{Z}_t=\frac{t}{1+t^2}\,Z_t,
\end{equation}
where \(\smash{\vb{X}_t=R_t\big(\kern-0.1em\cos\Phi_t\sm\hat{\vb{x}}+\sin\Phi_t\sm\hat{\vb{y}}\big)+Z_t\sm\hat{\vb{z}}}\) is the position of the particle at time \(\smash{t}\). For an arbitrary initial condition \(\smash{\vb{X}_0}\), the solutions are given by,
\begin{equation}
    R_t=R_0,\quad \Phi_t=\Phi_0+ t,\quad \text{and}\quad Z_t=Z_0\sqrt{1+t^2}.
\end{equation}
Physically, the Bohmian trajectories are circular helices of radius \(\smash{R_0}\) that circulate in an anticlockwise sense about the magnetic field axis, as do the classical trajectories governed by the Lorentz force\footnote{A difference, however, is that the velocity component parallel to the magnetic field remains constant in classical mechanics, while that in Bohmian mechanics is {\emph {asymptotic}} to a constant, as \(\smash{t\to\infty}\). Note that the Bohmian dynamics is inherently non-Newtonian, so the similarities between the Bohmian and Newtonian trajectories found in this example are more of an exception than a rule.}. The angular velocity is a constant of the motion, given by \(\smash{qB_0/ m}\) (\(\smash{=1}\) in our units), as is the radial coordinate \(\smash{R_t}\).

All trajectories starting with \(\smash{0<Z_0<L}\) propagate rightward, striking \(\smash{z=L}\) at time
\begin{equation}
    \tau(\vb{X}_0) = \sqrt{(L/Z_0)^2-1},
\end{equation}
while other trajectories do not cross \(\smash{L}\) at all, contributing to the non-detection probability \(\smash{P(\infty)}\). Now, recalling that the initial positions realized in a sequence of experiments are \(\smash{\rho_0=|\psi_0|^2}\)--distributed, we have
\begin{align}
    \Pi_{\text{BM}}(\tau) &= \int_0^{2\pi}\!\!\!d\Phi_0\!\int_0^{\infty}\!\!\!dR_0~R_0\!\int_0^L\!\!dZ_0~\rho_0(\vb{X}_0)\nonumber\\
    &\kern3cm\times\,\delta\big(\tau-\tau(\vb{X}_0)\big),
\end{align}
the Bohmian arrival-time probability density, associated with the non-detection probability
\begin{align}
    P_{\text{BM}}(\infty)&= \int_0^{2\pi}\!\!\!d\Phi_0\!\int_0^{\infty}\!\!\!dR_0~R_0\!\int_{-\infty}^{\infty}\!\!dZ_0~\rho_0(\vb{X}_0)\sm\big[\theta(-Z_0)\nonumber\\[2pt]
    &\kern3.5cm\,+\,\theta(Z_0-L)\big].
\end{align}
The latter is the \(\smash{\rho_0}\) measure of precisely those initial conditions for which the particle does not reach \(\smash{z=L}\) at any finite \(\smash{t}\).~In the present example, \(\smash{\rho_0(\vb{X}_0)=\pi^{\sm-\,3/2}}\) \(\smash{\exp(-\sm R_0^2-Z_0^2)}\), for which the integrals over \(\smash{R_0}\) and \(\smash{\Phi_0}\) can be easily evaluated, yielding
\begin{align}\label{nicely}
    \Pi_{\text{BM}}(\tau)= \frac{1}{\sqrt{\pi}}\kern-0.1em\int_0^L\!\!\!dZ_0~\delta\!\left(\tau-\sqrt{(L/Z_0)^2-1}\sm\right)\exp(-Z_0^2),
\end{align}
and
\begin{align}\label{pinfBM}
    P_{\text{BM}}(\infty)&= \frac{1}{\sqrt{\pi}}\left(\int_{-\infty}^0\!+\int_L^{\infty}\right)dZ_0~\exp(-Z_0^2)\nonumber\\[3pt]
    &= 1-(1/2)\,\text{erf}\sm(L),
\end{align}
where \(\smash{\text{erf}\sm(\cdot)}\) is the error function. For \(\smash{L\gg1}\), \(\smash{\text{erf}\sm(L)\approx1}\), hence \(\smash{P_{\text{BM}}(\infty)\approx 1/2}\), which substantiates our heuristic remark under \eqref{qfabove}. Finally, substituting \(\smash{Z_0=L/\sqrt{\xi^2+1}}\) in \eqref{nicely}, we have
\begin{align}
    \Pi_{\text{BM}}(\tau) &= \frac{L}{\sqrt{\pi}}\int_0^{\infty}\!\!\!d\xi~\frac{\xi}{(1+\xi^2)^{3/2}}\,\delta(\tau-\xi)= \theta(\tau)\Pi_{\text{QF}}(\tau),
\end{align}
which along with \eqref{pinfBM} is correctly normalized \`a la Eq. \eqref{normal}. The quantum flux distribution is duly reproduced in the end since every Bohmian trajectory in this case crosses \(\smash{z=L}\) \emph{at most once}, fulfilling \eqref{preCPC}, equivalently \eqref{CPC}, as noted below \eqref{Jz}.
\section{Conclusion}\label{Conc}
The ABK result has been used by many authors to support and promote acceptance of the time-energy uncertainty relation. Our considerations suggest serious reconsideration of these ambitions, once the unphysical ``negative arrival times'' inherent in them are taken seriously. The QF distribution is a natural candidate for the arrival-time distribution in quantum mechanics but does not enjoy unrestricted applicability due to the backflow effect. This defect is remedied by the Bohmian arrival-time distribution from which the QF was derived. On the other hand, there are often serious difficulties met in realizing stable backflow situations experimentally (see, however, \cite{MugaBEC,MugaBEC1,DD}). Thus, \(\Pi_{\text{QF}}\) is unproblematic for most practical purposes, even though it does not follow from a generalized quantum observable (or POVM)\footnote{From a Bohmian perspective, this is not a showstopper since the notion of an observable is absent in the fundamental posits of this theory (see also \cite{DGZOperators}). In this context, one could even argue ``that the significance of the [observables] has been exaggerated, in the sense that elements entering into useful mathematical techniques have been raised to the level of fundamental concepts entering into the physical theory'' \cite{BohmT}.}.

Finally, the ``standard distribution'', understood as a natural generalization of the ABK free-motion result, cannot be applied to vector potentials in the manner explored in Sec. \ref{Bfield}, as it fails to be gauge-invariant. Finding a gauge-invariant generalization therefore remains a challenging and important task for its proponents, as electromagnetic fields are essential to any realization of a TOF experiment. As a first step, one might restrict attention to freely moving charged particles alone, where free motion is taken to mean, as usual, vanishing \(\smash{\vb{E}}\) and \(\smash{\vb{B}}\). The electromagnetic potentials, however, are non-zero, satisfying \(\smash{\vb{A} = \grad{\lambda}}\) and \(\smash{V = -\,\partial \lambda/\partial t}\), where \(\smash{\lambda(\vb{x},t)}\) is an arbitrary real function. The most general Hamiltonian describing this motion is thus \(\smash{\hat{H}(\lambda)=\big(\hat{\vb{p}}-q\sm\grad\lambda\big)^2\kern-0.1em/(2m)-q\sm\partial\lambda/\partial t}\).~A self-consistent free-motion TOF distribution, \(\smash{\Pi_\lambda(\tau)=F\big(\psi^{(\lambda)}_\tau,\lambda\big)}\), given by some positive functional \(\smash{F}\), and a solution \(\smash{\psi^{(\lambda)}_\tau}(\vb{x})\) of Schr\"odinger's equation with Hamiltonian \(\smash{\hat{H}(\lambda)}\), would be expected to have a vanishing functional derivative, i.e., \(\smash{\delta \Pi_\lambda(\tau)/\delta\lambda =0}\).~In addition, one could require that \(\smash{\Pi_\lambda(\tau)}\) reproduce \(\smash{\Pi_{\text{Kij}}(\tau)}\) whenever \(\smash{\lambda(\vb{x},t)}\) is a constant, for which \(\smash{\hat{H}(\lambda)=\hat{\vb{p}}^2/(2m)}\). It is an interesting question whether such a \(\smash{\Pi_\lambda}\) could be associated with a POVM and, if so, what general POVM structures would be compatible with it.

\section*{Acknowledgements}
We are grateful to J. M. Wilkes and W. Struyve for carefully reviewing the manuscript. The interesting \(\smash{B_0\to0}\) limit of Sec. \ref{Bfield} was suggested by Anirudh Chandrasekaran. M. N. acknowledges funding from the Elite Network of Bavaria, through the Junior Research Group `Interaction Between Light and Matter'.
\bibliography{FluxvsKij}
\end{document}